\documentclass[graybox]{svmult}

\usepackage{mathptmx}       
\usepackage{amssymb}
\usepackage{amsmath}
\usepackage{graphicx}
\usepackage{subfigure}

\usepackage{helvet}         
\usepackage{courier}        
\usepackage{type1cm}        
\usepackage{makeidx}         
\usepackage{multicol}        
\usepackage[bottom]{footmisc}

\makeindex

\usepackage{diagbox}
\usepackage{color}
\usepackage{changes}

\newcommand{\A}{\textit{A }}
\newcommand{\B}{\textit{B }}

\begin{document}

\title*{Prepaid or Postpaid? That is the question.\\ Novel Methods of Subscription Type Prediction in Mobile Phone Services}

\titlerunning{Prepaid or Postpaid? That is the question}

\author{Yongjun Liao,  Wei Du, M\'{a}rton Karsai, Carlos Sarraute, Martin Minnoni and Eric Fleury}
\institute{Yongjun Liao (\email{yongjun.liao@inria.fr}), \\ M\'{a}rton Karsai (\email{marton.karsai@inria.fr}),\\ Eric Fleury (\email{eric.fleury@inria.fr}) \at Univ Lyon, ENS de Lyon, Inria, CNRS, UCB Lyon 1, LIP UMR 5668, IXXI, F-69342, Lyon, France.
\and Wei Du (\email{wei.du@insa-lyon.fr}) \at Univ Lyon, INSA Lyon, Inria, CITI, F-69621 Villeurbanne, France.
\and Carlos Sarraute (\email{charles@grandata.com}), Martin Minnoni, (\email{martin@grandata.com}) \at Grandata Labs, Bartolome Cruz 1818 V. Lopez. Buenos Aires, Argentina.
}



%

\maketitle

\abstract{In this paper we investigate the behavioural differences between mobile phone customers with prepaid and postpaid subscriptions. Our study reveals that (a) postpaid customers are more active in terms of service usage and (b) there are strong structural correlations in the mobile phone call network as connections between customers of the same subscription type are much more frequent than those between customers of different subscription types. Based on these observations we provide methods to detect the subscription type of customers by using information about their personal call statistics, and also their egocentric networks simultaneously. The key of our first approach is to cast this classification problem as a problem of graph labelling, which can be solved by max-flow min-cut algorithms. Our experiments show that, by using both user attributes and relationships, the proposed graph labelling approach is able to achieve a classification accuracy of $\sim 87\%$, which outperforms by $\sim 7\%$ supervised learning methods using only user attributes. In our second problem we aim to infer the subscription type of customers of external operators. We propose via approximate methods to solve this problem by using node attributes, and a two-ways indirect inference method based on observed homophilic structural correlations. Our results have straightforward applications in behavioural prediction and personal marketing.}

\section{Introduction}
\label{sec:introduction}

In most of the countries mobile phone operators propose two subscription options for their customers. In case of \emph{prepaid} subscription, credit is purchased in advance by the customer and access to the service is granted only if there is available credit. On the contrary, in case of \emph{postpaid} subscription, a user is engaged in a long-term contract with the operator, and service is billed according to the usage at the end of each month. Typically, a contract specifies a limit or ``allowance'' of minutes and text messages for what a user is billed at a flat rate, while any further usage incurs extra charges. Due to these differences in the level and time of engagement, different options of subscriptions may be adopted by typical user groups characterised by similar age, location, socioeconomic status, etc., Moreover, due to effects of homophily and social influence these people may be even connected to each other and communicate frequently thus forming a locally homogeneous sub-structures in the larger mobile communication network. Such correlations between the communication dynamics, structure and customer features can potentially be used to differentiate between customers.

In this paper, we are interested in such behavioural differences of prepaid and postpaid customers for at least three reasons.
\begin{itemize}
 \item The detection of typical patterns of service usage corresponding to different subscription types may provide useful information for the operator when planning network management and service pricing.
 \item The identification of customers with atypical service usage patterns of their actual subscription type may help the design of better direct advertising and personal services.
 \item The inference of the subscription type of customers of other operators could help direct marketing to convince customers to migrate to the actual provider.
\end{itemize}
Our study based on the analysis of customers of a single operator reveals that the main difference between prepaid and postpaid users is in the way they use mobile phone services. As a first result we found that postpaid users are more active and on average make more calls, towards more people as compared to prepaid users. We also found that only relying on attributes of call statistics, such as total duration and number of outgoing calls, we can classify fairly well customers, with an overall accuracy of $\sim 80\%$, by using standard machine learning tools. 

However, to achieve better results, further we investigated the proxy social network of users mapped out from their mobile phone communication events. This analysis revealed strong structural correlations among users of the same subscription type, who actually called each other $\sim 3$ times more frequently than others holding the opposite subscription type. This observation suggested that the classification accuracy could potentially be improved by algorithms considering not only individual activity patterns but also relationships between users at the same time. This is important, as, on the one hand, classification methods typically focus on finding distinct attributes associated to each class, while ignoring the network topology; and on the other hand, methods aiming to partition networks (e.g. tools of community detection) typically ignore attributes associated to each node. By considering simultaneously the two sources of information, we propose an approach to combine classification and graph partitioning techniques in a unified framework To this end, we formulate the classification problem as a problem of graph labelling, i.e. assigning a label, either prepaid or postpaid, to each user by taking into account both the user attributes and the structure in the mobile phone call network. This graph labelling problem can be solved by algorithms such as max-flow min-cut~\cite{maxflowmincut,graphcuts}, which in turn lead to an improved classification accuracy of $87\%$.

Inspired by these results, we further addressed the practical problem of inferring the service type of customers of external operators from the view of the host provider. Such information is important for mobile service operators to expand business in the already crowded telecommunication market. In this settings we consider two sets of customers: one set includes customers of the host operator (\textit{company users}) with the subscription type and all call records available. The other set includes customers of external operators (\textit{non-company users}) with no information available but only about their interactions with company users. As mobile service operators never share the data of their customers for the concerns of economy, security and privacy, an operator can only perform this inference problem by using visible data such as the subscription types and the communication records of its own company users (including calls with non-company users) but without knowing anything about the communications between any two non-company users. However, by assuming that the same structural correlations exists between customers of different providers as observed between company users, we propose possible ways to infer and indirectly verify subscription types of non-company users, even without knowing their real values. This explanatory study below shows that our direct inference method, based on node attributes and structure extracted from call data between two operators, alone can achieve an accuracy of $\sim 70\%$, while indirect inference methods may also provide meaningful insights.

The rest of the paper is organised as follows. In Section \ref{sec:relatedwork} we discuss the related works, in Section \ref{sec:dataset} and \ref{sec:classification} we describe the dataset in use and the classification results using machine learning methods based merely on node attributes. Section \ref{sec:labeling} presents our graph labelling algorithm which exploits both node attributes and the local network structure. In Section \ref{sec:inferfromoutside} we introduce our methods of inference of the subscription types of non-company users. Finally in Section \ref{sec:conclusions} we conclude our work and discuss some potential future directions of research.

\section{Related Work}
\label{sec:relatedwork}

Our work is motivated by the work of Yang et al.~\cite{communitydetectionwithnodeattributes} in which network communities are identified by using both user attributes and the structure of the social network. However, one important distinction of the present work is that in~\cite{communitydetectionwithnodeattributes} the authors aimed to cluster nodes with similar attributes into some apriori unknown number of communities, whereas we aim to find discriminating attributes of nodes and assign them to two pre-defined communities. In addition, our approach is based on graph labelling and solved by simple graph algorithms, whereas~\cite{communitydetectionwithnodeattributes} is based on a generative model optimised by block-coordinate descent. Other related works on community detection with user attributes include~\cite{Burton:2014:DSC:2663597.2641759,Sun:2012:RSC:2140436.2140437,Xu:2012:MAA:2213836.2213894}. However, all of them focus on clustering nodes in a network into priori unknown partitions or communities, while none of them address the classification of nodes into pre-defined classes as here. A related classification problem which has been well studied in the literature is the inference of user demographics such as nationality, gender and age from social and mobile call networks~\cite{Dong:2014:IUD:2623330.2623703,DBLP:conf/icwsm/ChenWAW15,DBLP:journals/corr/MalmiW16,inferring-the-demographics-of-search-users,classificationbyfactograph}. While interesting, most of the previous work exploited only node attributes some of which are computed from local connections.

Knowledge transfer from one network to another is also a hot topic in the field of machine learning and social network analysis. Most of these works focused on the inference of the structure, i.e. link prediction, using information from multiple networks. For example, Eagle et al. inferred the friendship network from mobile call data~\cite{citeulike:5455141} and Dong et al.~\cite{Dong:2012:LPR:2471881.2472671} and Ahmad et al.~\cite{5693393} predicted links across heterogeneous networks, which may only be partially visible. In addition, Tang et al.~\cite{Tang:2012:IST:2124295.2124382} studied the problems of relationship classification across networks and Kong et al.~\cite{Kong:2013:IAL:2505515.2505531} inferred the anchor links between users in different social networks, i.e. the same user with different accounts. In contrast to most previous work, our study on cross-network inference focuses on the classification of nodes in the targeted network using only the connections between two networks.

\section{Dataset}
\label{sec:dataset}

Our dataset is a sequence of anonymised Call Detailed Records (CDR) collected by a single operator in a Latin American country. It contains $\sim 280$ millions of call and SMS events recorded during one month between $\sim 20$ millions of mobile phone users. Each CDR records the starting time of the event (date and time), communication type (call or SMS), the call duration (in case of call event), the anonymised identifier of the user originating the call (caller), and the anonymised identifier of the person receiving the call (callee), but not the content of that transaction. Note that from the caller and callee at least one is a customer of the operator.

For the first part of our study we constructed a dataset (DS1) of $3.2$ million users, who are actual customers of the operator, and have available information about there subscription type being prepaid or postpaid. After filtering out users with either too few (total duration of calls $<10$ seconds) or unrealistically many (total duration of calls $>100,000$ seconds) communications, we obtained about $1.3$ million prepaid and $1.2$ million postpaid users in total. For the purpose of our investigation, from the $190$ million CDRs of $2.5$ million selected users, we constructed a directed weighted communication network with $90$ million links between interacting users. More precisely, the communication network $G=(V,E)$ was define by mobile users as nodes connected by directed weighted links. Links $(u,v,w)\in E$ between users $u$ and $v$ (where $u,v\in V$) were drawn if at least one communication event took place between them during the observation period. The weight $w$ of a directed link was defined as the number of communication events or the total duration of calls took place between $u$ and $v$ (actual weight definition is specified later). Note that DS1 is balanced with roughly the same number of users for each type. However, our approach does not rely on this fact as the only assumption we take is based on a structural correlation in the communication network, which we will verify in the next sections.

For the second part of our study we considered the whole available communication sequence to construct a dataset (DS2). After similar filtering as before, we obtained $3.1$ million company users ($1.6$ million prepaid and $1.5$ million postpaid), with full list of CDRs available, and $14.1$ million non-company users, with CDRs available only with company users. Note that the increase of company users in DS2 as compared to DS1 is due to the company users who only interacted with non-company users, thus appear only in DS2 with non-zero degree. In DS2 in total we had $190$ million calls between company users and $90$ million calls between company and non-company users. Using these events we constructed a network, which contained $90$ million links between company users (just as in DS1), and $44.6$ million links between company and non-company users.

\section{User Attributes and Classification}
\label{sec:classification}

\begin{figure*}[t]
    \centering
     \subfigure[]{\includegraphics[width=0.46\linewidth]{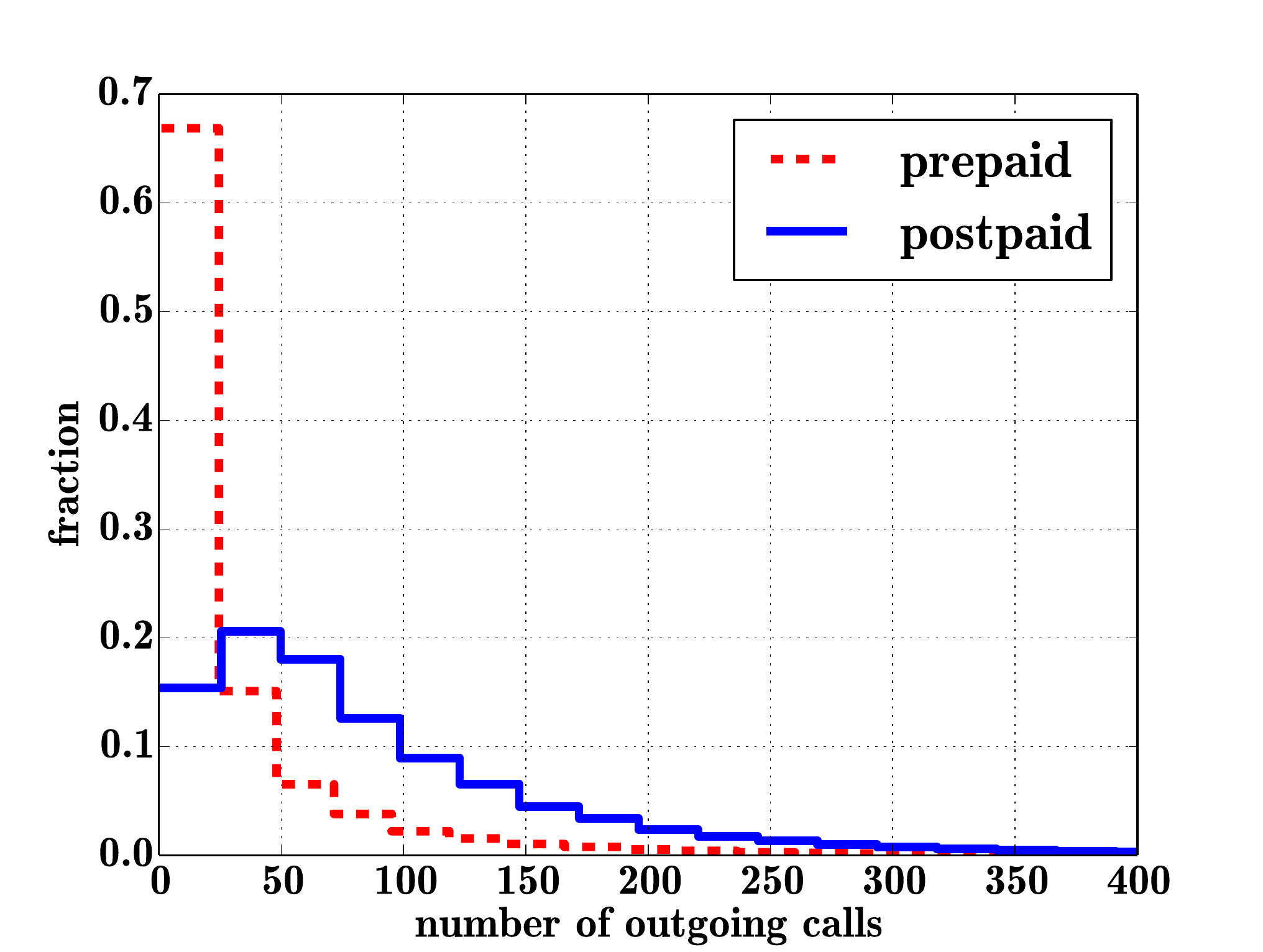}}
     \qquad
     \subfigure[]{\includegraphics[width=0.46\linewidth]{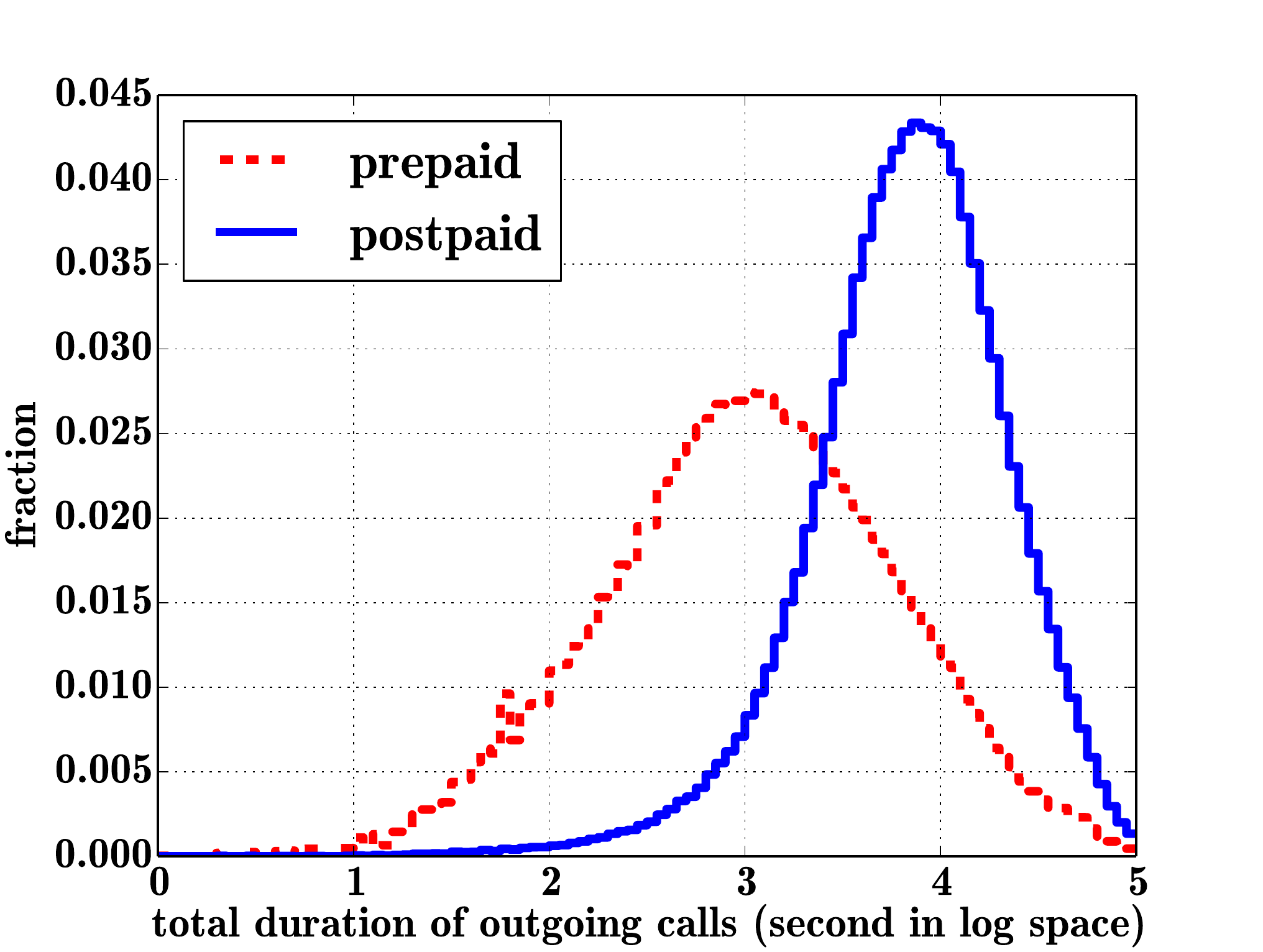}}\\
     \subfigure[]{\includegraphics[width=0.46\linewidth]{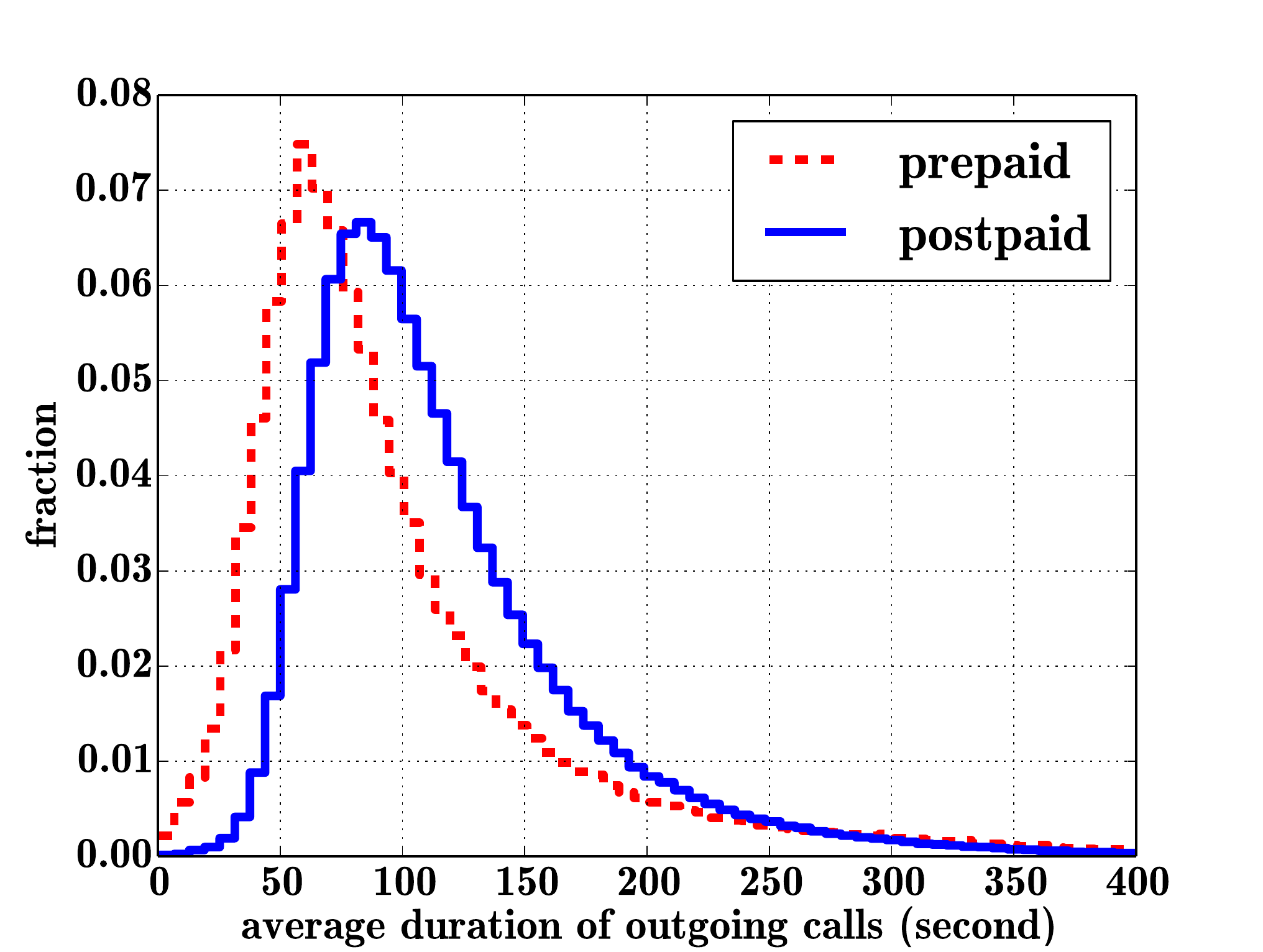}}
     \qquad
     \subfigure[]{\includegraphics[width=0.46\linewidth]{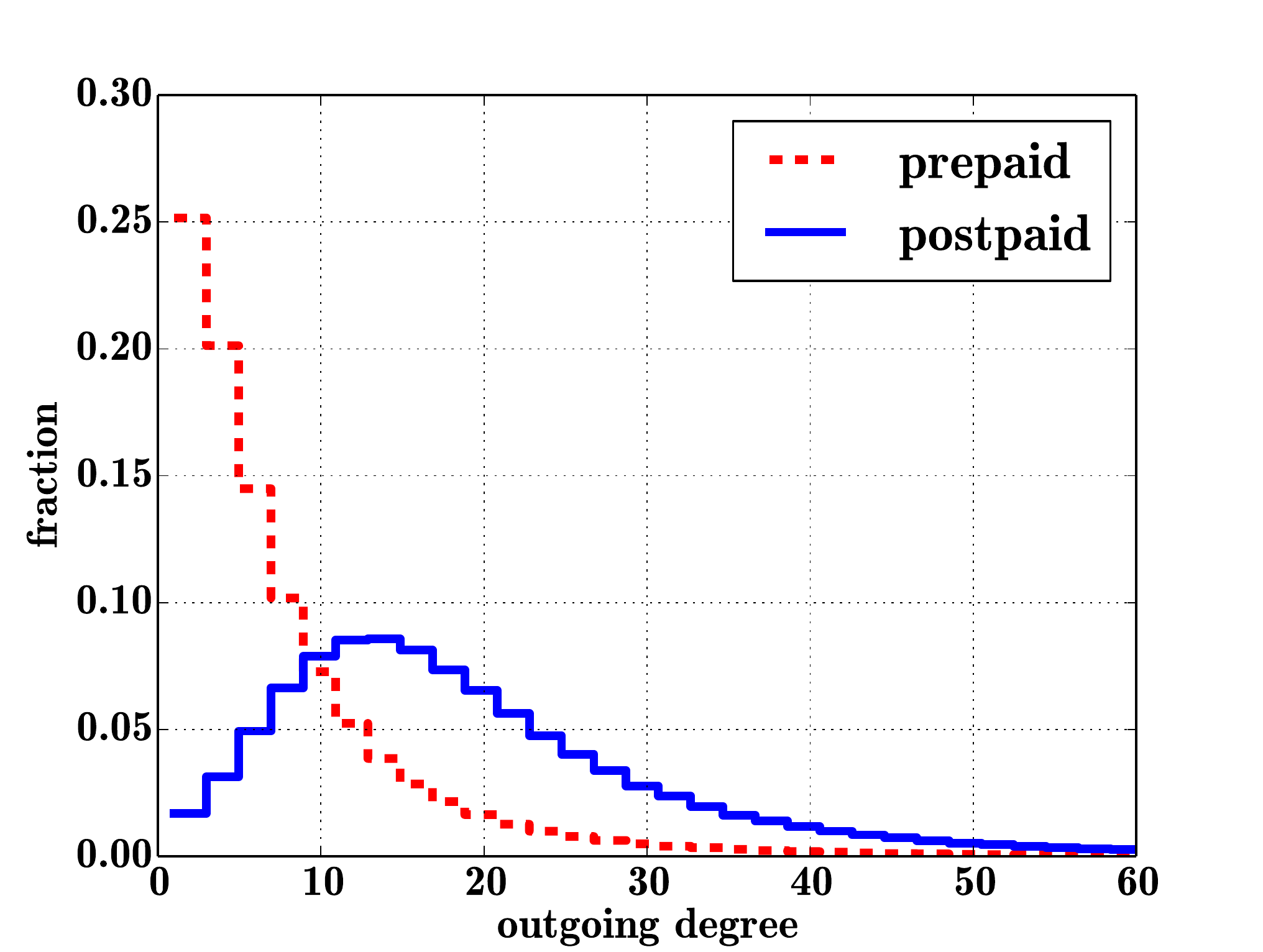}}
 \caption{Distribution of user call activity attributes as (a) total number of outgoing calls, (b) the total (in the logarithmic space) and (c) average duration of each outgoing call and (d) the outgoing degree of each user.}
 \label{fig:attributedistribution}
\end{figure*}

For each user, using the first dataset DS1 we extracted a number of attributes related to call statistics including:
\begin{itemize}
 \item total number of outgoing calls;
 \item total duration of outgoing calls;
 \item mean and standard derivation of the duration of outgoing calls;
 \item outgoing degree (number of callees) in the CDR network.
\end{itemize}
These attributes turned out to be very different for prepaid and for postpaid users as shown in Figure \ref{fig:attributedistribution} where the distributions of (a) the total number of outgoing calls, (b) the total and (c) the average duration of outgoing calls, and (d) the outgoing degrees are depicted. From these measures we found that on average postpaid users make $2.9$ times more calls to $2.5$ times more people as compared to prepaid users. Note that we tested and found that attributes related to incoming calls and SMS are not informative. By using these attributes, we built our classifiers with different machine learning methods including support vector machines (SVM), Boosting (AdaBoost), Naive Bayes and Decision Tree to estimate the subscription type of each user. In every experiment presented in the paper, the classifiers were trained using $20,000$ randomly sampled users, half of them prepaid and half postpaid, and tested on the rest of the $2.5$ million users. After these experiments we found that SVM and AdaBoost achieved the best accuracy of about $80\%$, while Naive Bayes achieved about $77\%$ and Decision Tree about $71\%$. The confusion matrix (each column of the matrix represents the instances in a predicted class while each row represents the instances in an actual class) for SVM is given in Table \ref{tab:confusion_classification}.

Note that the Naive Bayes method is a probabilistic learning method providing a probability for each node to belong to one of the subscription type, unlike SVM and AdaBoost, which provide binary values for labelling. For this reason, although less accurate, Naive Bayes is used later to define our graph labelling approach.

\begin{table}[ht]
\caption{Confusion matrix by SVM.}
\centering
\begin{tabular}{|p{3.6cm}|p{3.6cm}|p{3.6cm}|}
\hline accuracy=\textbf{0.803}&{postpaid}&{prepaid}\\\hline
{postpaid}&\textbf{0.822}&0.178\\\hline
{prepaid}&0.214&\textbf{0.786}\\\hline
\end{tabular}
\label{tab:confusion_classification}
\end{table}

Besides the above-mentioned attributes, which are related only to call statistics, attributes extracted from the social-communication network may also help our prediction. The analysis of the network revealed a strong homophilic structural correlation, as calls between the same type of users (prepaid or postpaid) appeared on average $3$ times more often than the ones across the two types of users. The matrix disclosing the fraction of calls by the two type of users is given in Table \ref{tab:confusion_call}. This highly partitioned structure suggests that potentially the classification accuracy can be further improved by exploiting the observed sparse connectivity between the two user sets.
\begin{table}[ht]
\caption{Proportion of calls within and between the two types of users.}
\centering
\begin{tabular}{p{2.3cm}|p{4.2cm}|p{4.2cm}}
\hline \diagbox[width=8em]{caller}{callee}&{prepaid}&{postpaid}\\\hline
{prepaid}&\textbf{0.791}&0.209\\\hline
{postpaid}&0.173&\textbf{0.827}\\\hline
\end{tabular}
\label{tab:confusion_call}
\end{table}

To test our conjecture, we extracted postpaid portion attributes, by using the known subscription types of neighbouring users for each user. The portion attributes in focus were:
\begin{equation}
F_n^i=n^{po}_i/n_i, \hspace{.2in} F_c^i=c^{po}_i/c_i, \hspace{.2in} \mbox{and} \hspace{.2in} F_d^i=d^{po}_i / d_i,
\end{equation}
where $F_n^i$ (resp. $F_c^i$, and $F_d^i$) assigns the portion of the $n^{po}_i$ number of postpaid users in the callee set (resp. the $c^{po}_i$ number of calls, and the $d^{po}_i$ duration of calls to postpaid users) and the $n_i$ number of users in the callee set (resp. the $c_i$ number of calls, and the $d_i$ total duration of calls) of a user $i$. With these extra portion attributes we could largely improve the accuracy of SVM and AdaBoost from $\sim 80\%$ to $\sim 89\%$. The confusion matrix for SVM is given in Table \ref{tab:confusion_classification_portion}.

However, in real settings it is impossible to measure these portion attributes, as they require the knowledge of the service type of each user, which are to be estimated. This way the outcomes provided by these methods are only informal in terms of our original classification problem as they use information what are assumingly not available. On the other hand, these methods can be used to detect false positive cases, where a user was assigned with a subscription type, while having another type of contract. This information can be used for direct marketing to provide services to costumers, which might better fit their needs. In any case, the positive results of this initial analysis motivated us to exploit the network topology in our new classification method defined in the next section.

\begin{table}[ht]
\caption{Confusion matrix for SVM with the proportion attributes.}
\centering
\begin{tabular}{|p{3.6cm}|p{3.6cm}|p{3.6cm}|}
\hline accuracy=\textbf{0.891}&{postpaid}&{prepaid}\\\hline
{postpaid}&\textbf{0.897}&0.103\\\hline
{prepaid}&0.115&\textbf{0.885}\\\hline
\end{tabular}
\label{tab:confusion_classification_portion}
\end{table}

\section{Classification with User Attributes and Network Topology}
\label{sec:labeling}

\subsection{Graph Labeling}

The key idea here is to exploit both user attributes and the network structure to detect the service type of each user. This is to be done without any prior knowledge about the subscription types of users. To do so, we cast the classification problem as a problem of graph labelling and show that this graph labelling problem is equivalent to the classic network flow problem, namely max-flow min-cut. According to the max-flow min-cut theorem~\cite{maxflowmincut}, in a flow network, the maximum amount of flow passing from the source $s$ to the sink $t$ is equal to the minimum cut, i.e. the smallest total cost of the links, which if removed would disconnect the source $s$ from the sink $t$.

\begin{figure}[!ht]
 \centering
 \includegraphics[width=0.5\linewidth]{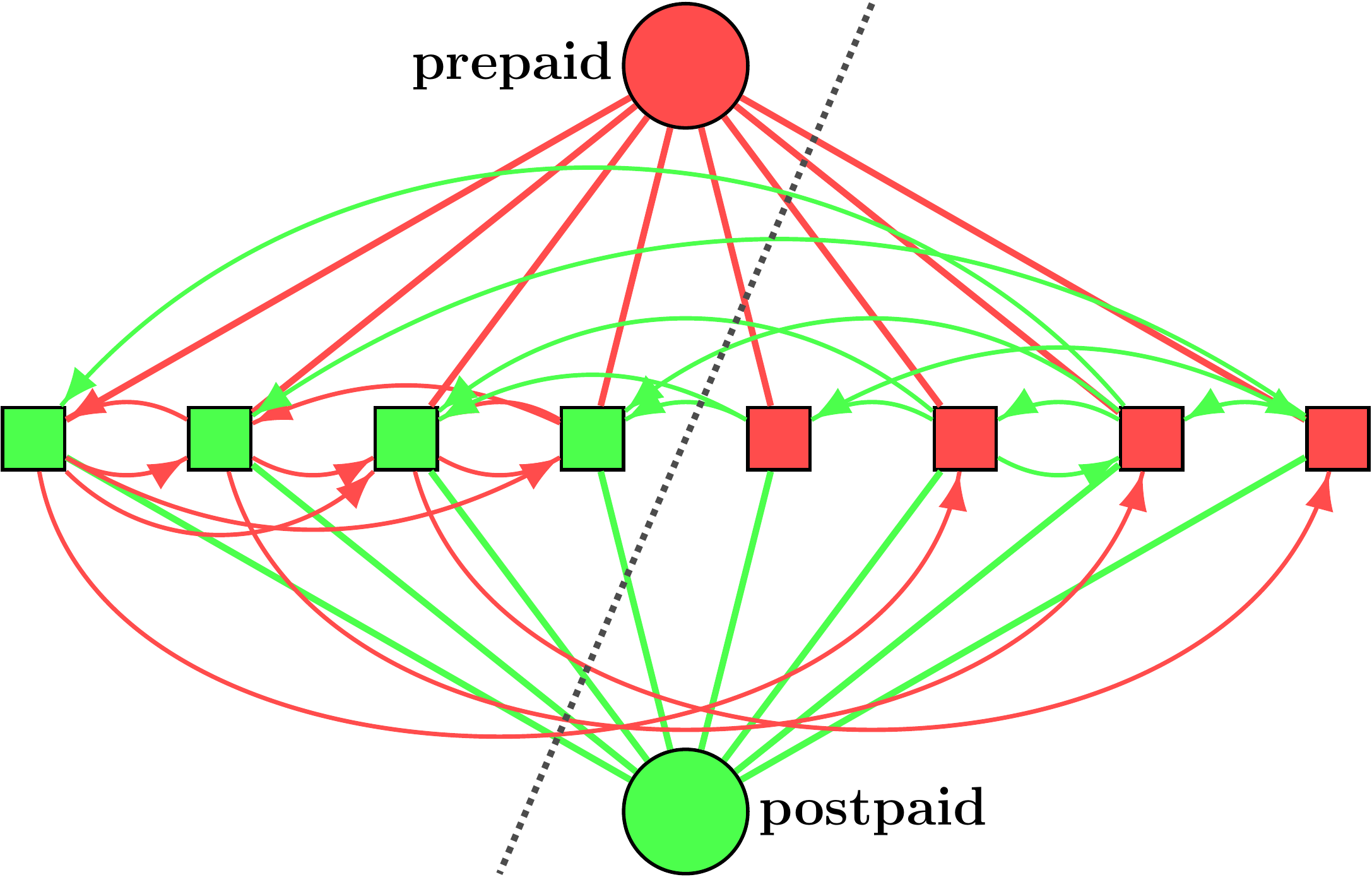}
 \caption{Graph labelling for the classification of the service type of each user. Square nodes represent users and circle nodes represent the labels of prepaid and postpaid. Each user is connected in the mobile call network and with both label nodes. A feasible cut will partition the graph so that no flow can go from the source (prepaid) to the sink (postpaid). The minimum cut (the dashed line) is the feasible cut with the least overall cost.}
 \label{fig:graphcuts}
\end{figure}

To see the connection between the two seemingly irrelevant problems, we take the CDR network $G=(V,E)$ and construct a new graph $G'=(V',E')$. We add two auxiliary nodes $V' = V \cup \{s,t\}$: this two nodes represents the two labels of prepaid and postpaid, which are assigned as the source $s$ and the sink $t$ respectively, as illustrated in Figure \ref{fig:graphcuts}. Initially, we connect each user to both label nodes: $E' = E \cup \{ (s, v) \,|\,\forall v \in V\} \cup \{(u,t)\,|\,\forall u\in V \}$. We call the links ($e=(u,v), u,v \in V$) between users as  \emph{social links} and the links ($e=(s,u)$ or $e=(u,t)$, $u \in V$) between any user and one of the two label nodes as the \emph{labelling links}. The costs associated to the social and labelling links will be described in the next section. However, independent of the definition of the cost functions, the following theorem guarantees that the minimum cut of the graph in Figure \ref{fig:graphcuts} assigns a label to each user.

\begin{theorem}
The minimum cut of a graph like in figure \ref{fig:graphcuts} will cut one and only one labelling link of the two labelling links of each user. 
\end{theorem}

\begin{figure}[h!]
 \centering
 \subfigure[Case 1]{\includegraphics[width=0.24\linewidth]{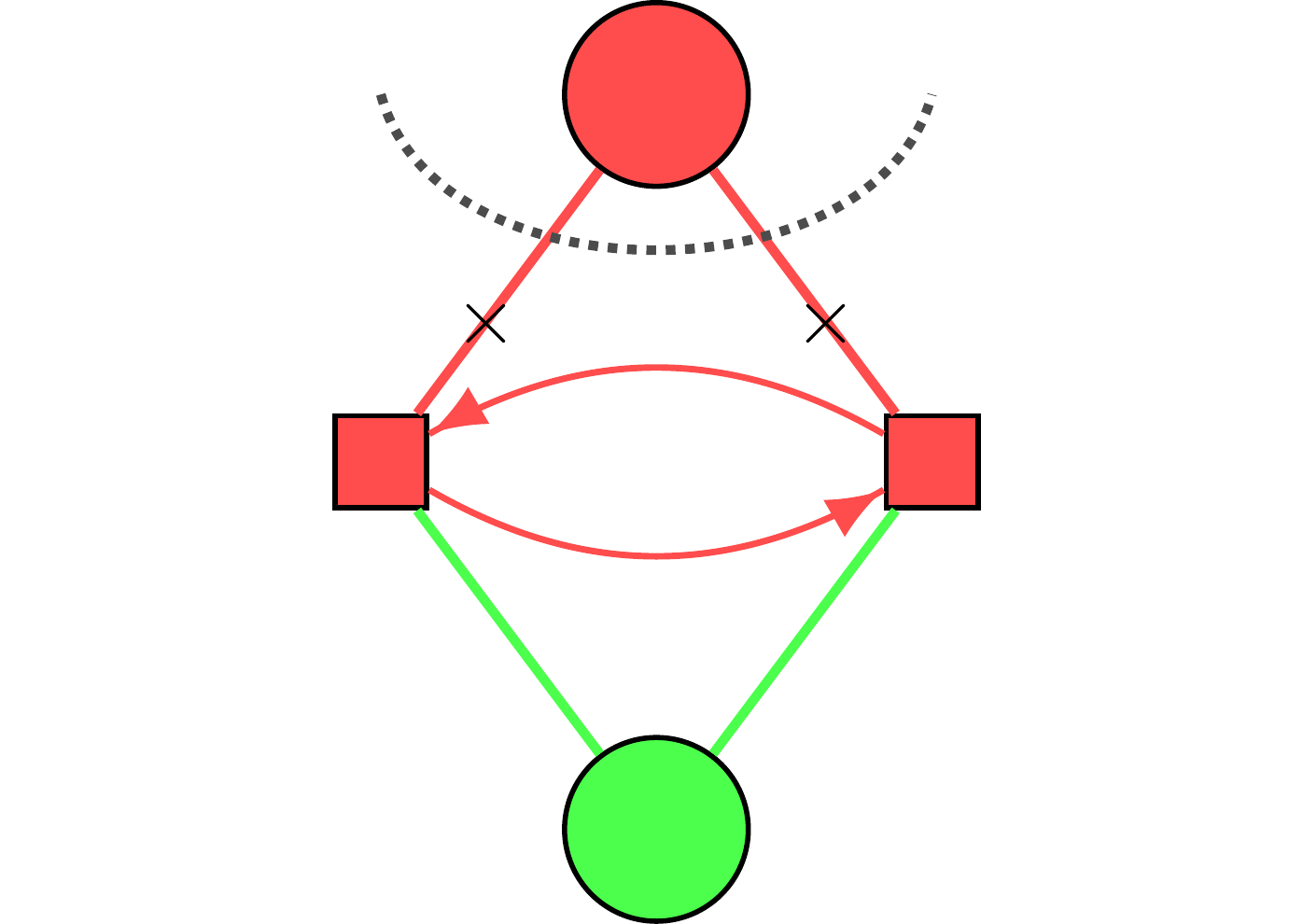}}
 \subfigure[Case 2]{\includegraphics[width=0.24\linewidth]{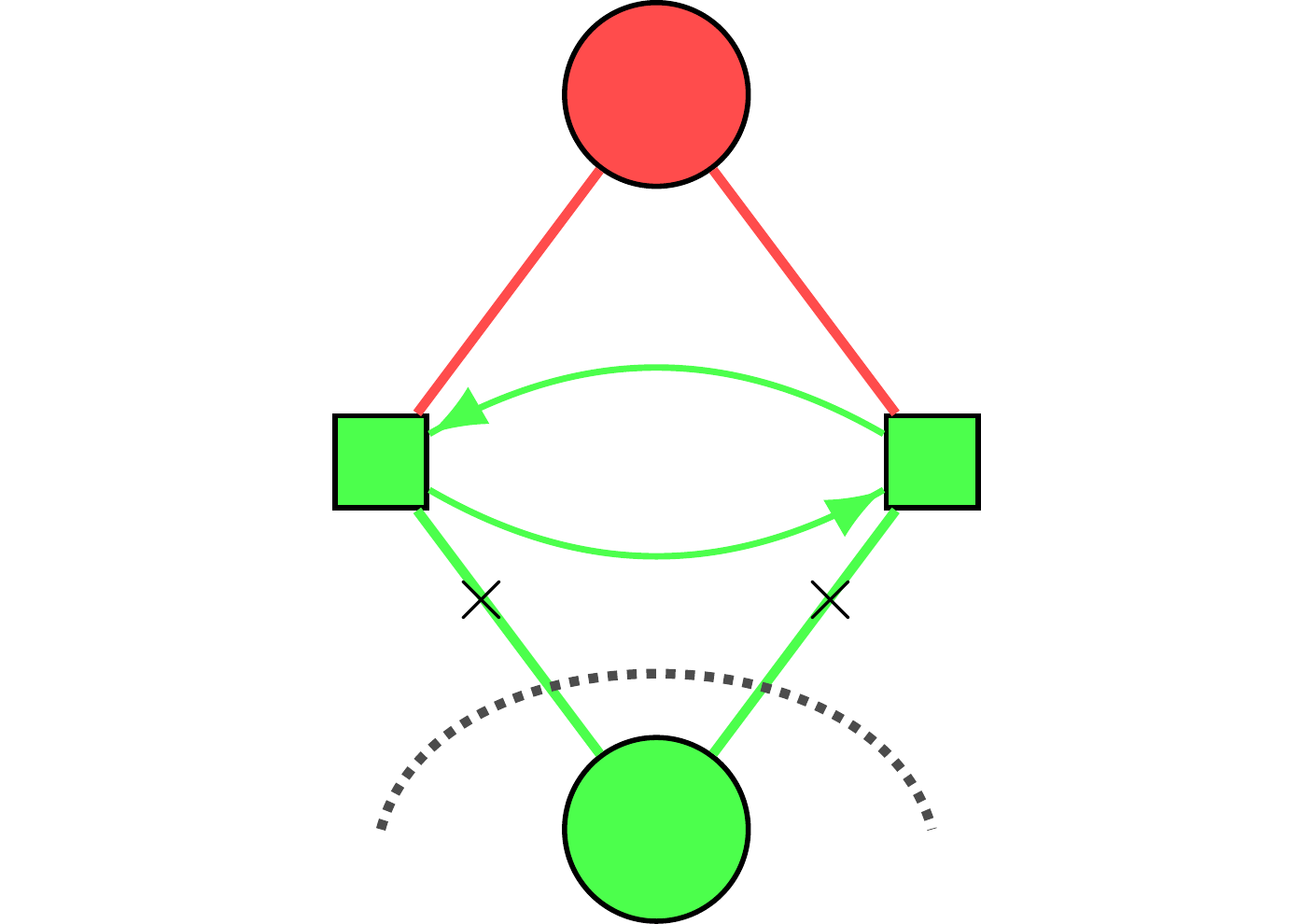}}
 \subfigure[Case 3]{\includegraphics[width=0.24\linewidth]{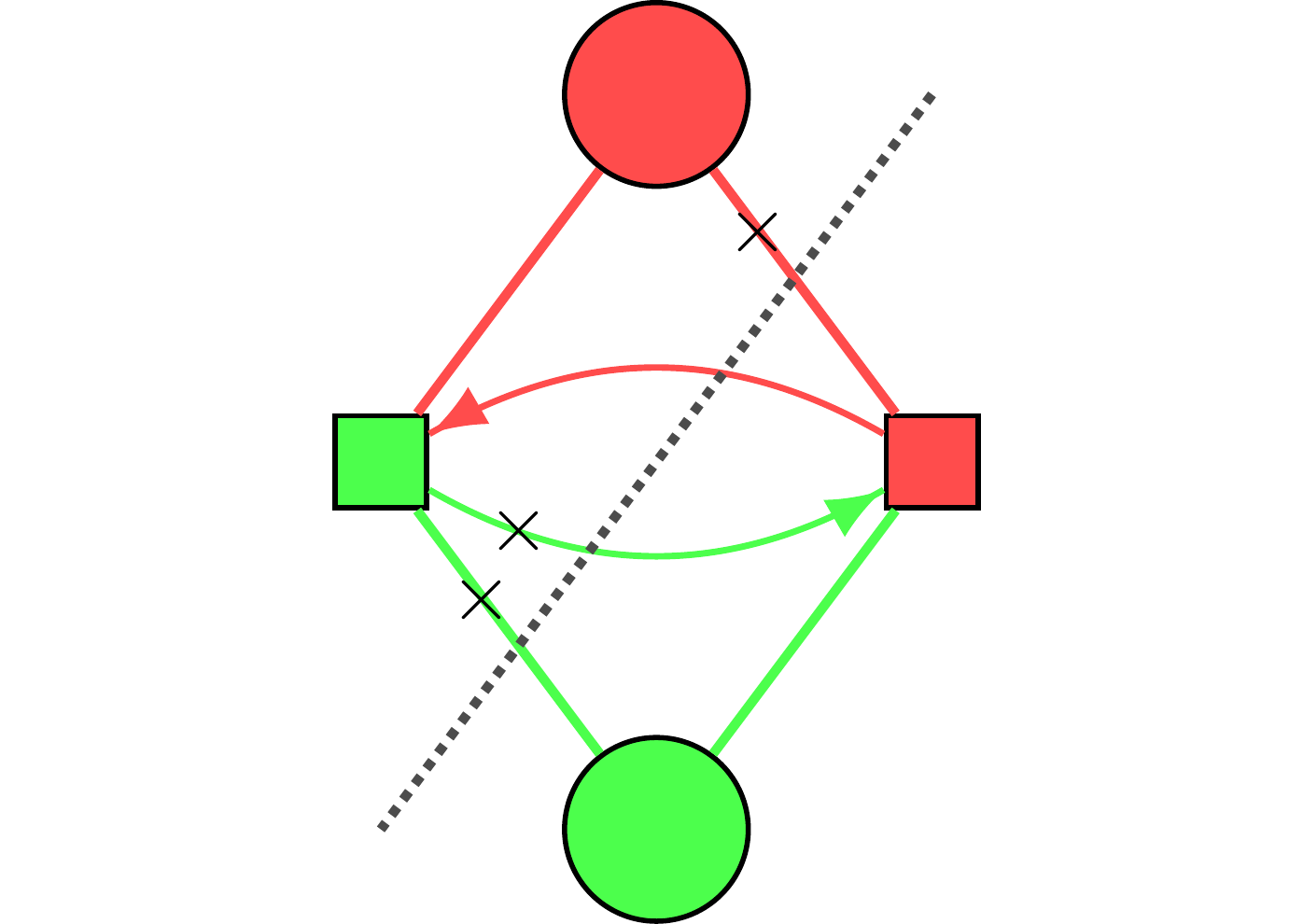}}
 \subfigure[Case 4]{\includegraphics[width=0.24\linewidth]{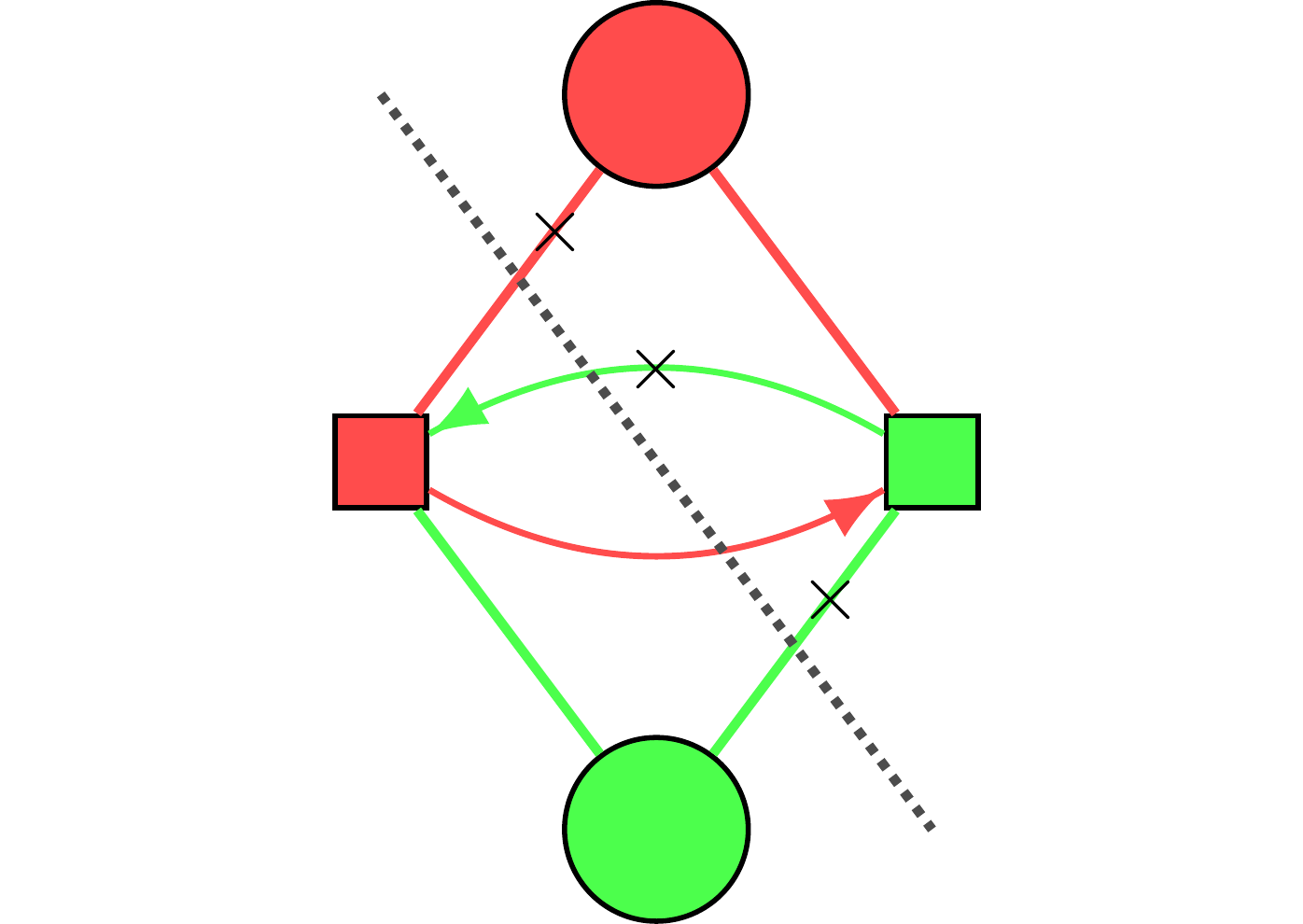}}
 \caption{4 possible cases for the feasible cut of the subgraph of a pair of connected users. Note that the labelling link which was cut indicates the labelling of a user. In the last 2 cases, only the one social link with a cross sign is cut.}
 \label{fig:subgraphcuts}
\end{figure}

\begin{proof}
By definition, a cut is feasible if it disconnects the source and the sink, and the minimum cut is the feasible cut with the least total cost. It is trivial that any feasible cut will cut at least one of the two labelling links of each user, however we need to show that any feasible cut that cuts both labelling links from a user is not a minimum cut of the graph. To see this, we consider a subgraph with a pair of doubly connected users. It is clear that the minimum cut of this subgraph can only be one of the four possible cases as shown in Figure \ref{fig:subgraphcuts}, all of which cut only one labelling link of each node. Which of these cuts apply in an actual subgraph depends on the cost defined on each link, but in any case, any graph containing such subgraphs must have the minimum cut which holds the same condition.
\end{proof}

It is important to note that the labelling link which is cut between a node and a label indicates the labelling of the node, which reflects the intuition that cut links tend to have smaller costs.

\subsection{Cost Functions and Problem Formulation}

In order to extend our classification problem as a graph labelling or max-flow min-cut problem, we need to associate a cost to each link in Figure \ref{fig:graphcuts}. To this end, we use the information about both user attributes and relationships encoded in the social structure. For user $u$, let $x_u$ be the user attribute vector, extracted from call statistics as discussed in Section \ref{sec:classification}, and $f_u$ be the possible label of either prepaid or postpaid, represented by 0 and 1 respectively. On the one hand, we use Naive Bayes to compute the probability $P(f_u|x_u)$ that user $u$ is a prepaid ($f_u$=$0$) and postpaid ($f_u$=$1$) subscriber. Note that Naive Bayes was adopted as it provides the probability of the classification rather than just a binary classification result. Subsequently, we define the $D_u$ cost of each labelling link based on the the probability function $P(f_u|x_u)$ as:
\begin{align}
D_u(f_u=0) = - \text{log} (P(f_u=0|x_u)).\nonumber\\
D_u(f_u=1) = - \text{log} (P(f_u=1|x_u)).
\label{eq:dataterm}
\end{align}
This definition assigns a small cost to a labelling link if the probability of a user taking the corresponding label is large. Note that in Naive Bayes, the normalisation condition holds, i.e., $P(f_u=0|x_u)+P(f_u=1|x_u)=1$, while we force $P(f_u|x_u)>1e-10$ to avoid zero values in the logarithm of Eq.\ref{eq:dataterm}.

On the other hand, to define the cost of a social link, we rely on our earlier observations that postpaid users tend to have $2.5$ times larger $k_{out}$ outgoing degrees as compared to prepaid customers, and that the two types of users are sparsely connected with each other. The outgoing degree $k_{\textrm{out}}(u)$ of each user $u$ is a simple structural attribute, which turned out to be discriminative here. Thus, we define the cost of a social link $w_{(u,v)}$ as a slightly modified Ising model~\cite{whatenergy}, i.e.,
\begin{align}
w_{(u,v)}(f_u=0,f_v=0) &= w_{(u,v)}(f_u=1,f_v=1) = 0,\nonumber\\
w_{(u,v)}(f_u=0,f_v=1) &= \frac{1}{k_{\textrm{out}}(v)},\nonumber\\
w_{(u,v)}(f_u=1,f_v=0) &= \frac{1}{k_{\textrm{out}}(u)}.
\label{eq:smoothnessterm}
\end{align}
This definition assigns zero cost to a link, which connects two users of the same label (corresponding to the first two cases in Figure \ref{fig:subgraphcuts} \emph(a) and \emph{(b)}).  Otherwise, when the label is different, a small cost is assigned that is inversely proportional to the outgoing degree of the user labeled as a postpaid customer (corresponding to the last two cases in Figure \ref{fig:subgraphcuts} \emph(c) and \emph{(d)}). Note that we also tried to define the cost as a measure of social strength between connected users, such as the number of outgoing calls or duration of outgoing calls, but obtained worse performance.

With the above defined cost functions for each social link $(u,v)$, the graph labelling problem in Figure \ref{fig:graphcuts} can be interpreted as an energy minimisation problem formulated as:
\begin{equation}
 \min \sum_{u\in V} D_u(f_u) + \lambda \sum_{u,v\in V} w_{(u,v)}(f_u,f_v).
 \label{eq:energyminimization}
\end{equation}

Interestingly,  Eq.~\ref{eq:energyminimization} can be associated to a class of discrete optimisation problems, which has been well studied in the field of computer vision and machine learning~\cite{graphcuts}. In this terminology, the first term $D_u$, called the data term, reflects the cost of the labelling link between $u$ and label $f_u$, while the second term is the smoothness term, reflecting the costs of social links between users $u$ and $v$, which encourages connected users to take the same labels. Note that $\lambda$ is a parameter that controls the trade off between the two terms and its impact on accuracy will be described in the next section. Also note that Eq.~\ref{eq:energyminimization} can be solved efficiently by any standard max-flow min-cut methods among which the push-relabel algorithm was chosen. In addition, we remark that the cost functions defined in Eq. \ref{eq:dataterm} and \ref{eq:smoothnessterm} do not require the labels of the nodes as inputs as they merely rely on probabilities determined by the Naive Bayes method using user attributes other than portion attributes.

\subsection{Implementation and Results}

Based on the above defined model formulation we can solve the graph labelling problem using standard max-flow min-cut algorithms~\cite{algorithmenergyminimization}. Our experiments showed that the graph labelling approach achieved an accuracy of $\sim 84\%$ which is better than using supervised learning on user attributes alone. Note that the parameter $\lambda$ has an impact on the classification accuracy. If $\lambda=0$, we ignore completely social relationships and graph labelling degenerates to Naive Bayes using user attributes, while if $\lambda=+\infty$, we ignore completely user attributes and enforce all nodes to take the same label, resulting in an accuracy of $\sim 50\%$ due to the similar sizes of two types of users in our dataset. Thus, we tuned $\lambda$ by cross validation to our dataset and found the best value to be $100$. Enlarging $\lambda$ by 10 times resulted in an accuracy of $\sim 74\%$, whereas shrinking $\lambda$ by 10 times gave us $\sim 80\%$.

\begin{table}[ht]
\caption{Confusion matrix by our graph labelling approach.}
\centering
\begin{tabular}{|p{3.6cm}|p{3.6cm}|p{3.6cm}|}
\hline accuracy=\textbf{0.868}&{postpaid}&{prepaid}\\\hline
{postpaid}&\textbf{0.901}&0.099\\\hline
{prepaid}&0.163&\textbf{0.837}\\\hline
\end{tabular}
\label{tab:confusion_labeling}
\end{table}

\begin{table}[ht]
\caption{Classification accuracy of different methods using only user attributes.}
\centering
\begin{tabular}{p{5.4cm}p{5.4cm}}
\hline method & accuracy\\\hline
graph labeling with graph pruning & 0.868\\
graph labeling  & 0.844\\
SVM with user attributes  & 0.803\\
Naive Bayes with user attributes & 0.770\\\hline
\end{tabular}
\label{tab:accuracy}
\end{table}

In addition, we tried to prune the graph by fixing the label of some users under two conditions: if the probability of a user belonging to a class was larger than a threshold $\tau_1$ and if the average probability of its neighbours belonging to the same class was larger than another threshold $\tau_2$. In our experiments, we set $\tau_1=85\%$ and $\tau_2=65\%$ which led to fixed labels for about $16\%$ of users. This way of pruning improved the accuracy of our prediction to $87\%$. The confusion matrix of this method is given in Table \ref{tab:confusion_labeling}, while Table \ref{tab:accuracy} summarises the accuracy achieved by all different methods in applied so far.

Note that the runtime of our approach is about 0.5 hour on a desktop with 32G memory and CPU Xeon of 1.90GHz which is not that slow considering the size of our graph. In particular, we adopted the push-relabel method to solve the max-flow min-cut problem, which has the computational complexity of $O(V^2E)$, where $V$ and $E$ are the number of nodes and edges respectively. Also note that, our graph labelling algorithm utilises the Naive Bayes for modelling the probability that a user is a prepaid or postpaid subscriber. This calculation requires $20,000$ randomly sampled training data, as mentioned earlier. However, in the optimisation, we assign labels to all $2.5M$ users including ones in the training set as well. The training samples have no impact to the results as they are less than $1\%$ of the total users and they are only used for learning the cost of labelling links.

\section{Subscription type inference of non-company users}
\label{sec:inferfromoutside}

In most of the countries typically there are several mobile service providers, who compete for customers on the same market. However a provider has access to user data (e.g. subscription type) of its own customers only, while information about the users of other providers are limited to the CDRs between the company and non-company customers, recorded for billing purposes. In this section we use such CDRs to introduce two methods to infer the subscription type of non-company users from a company point of view. Such knowledge is certainly valuable for a provider e.g. to design advertisement strategies to motivate the churning of non-company users.

However, this is a difficult problem as the direct verification of inferred subscription types of non-company users is not possible. To come around this problem, we build on the assumption that call feature patterns, which were shown to be relevant in our classification problem earlier (see Section \ref{sec:classification}), actually characterise all customers on the market, not only the company users. This way such features may provide predictive power when characterising calling patterns between company and non-company users, thus can be effectively used for inference and help the design of new validation methods. 

\subsection{Inference using call statistics}

\begin{figure}[t]
    \centering
     \subfigure[]{\includegraphics[width=0.46\linewidth]{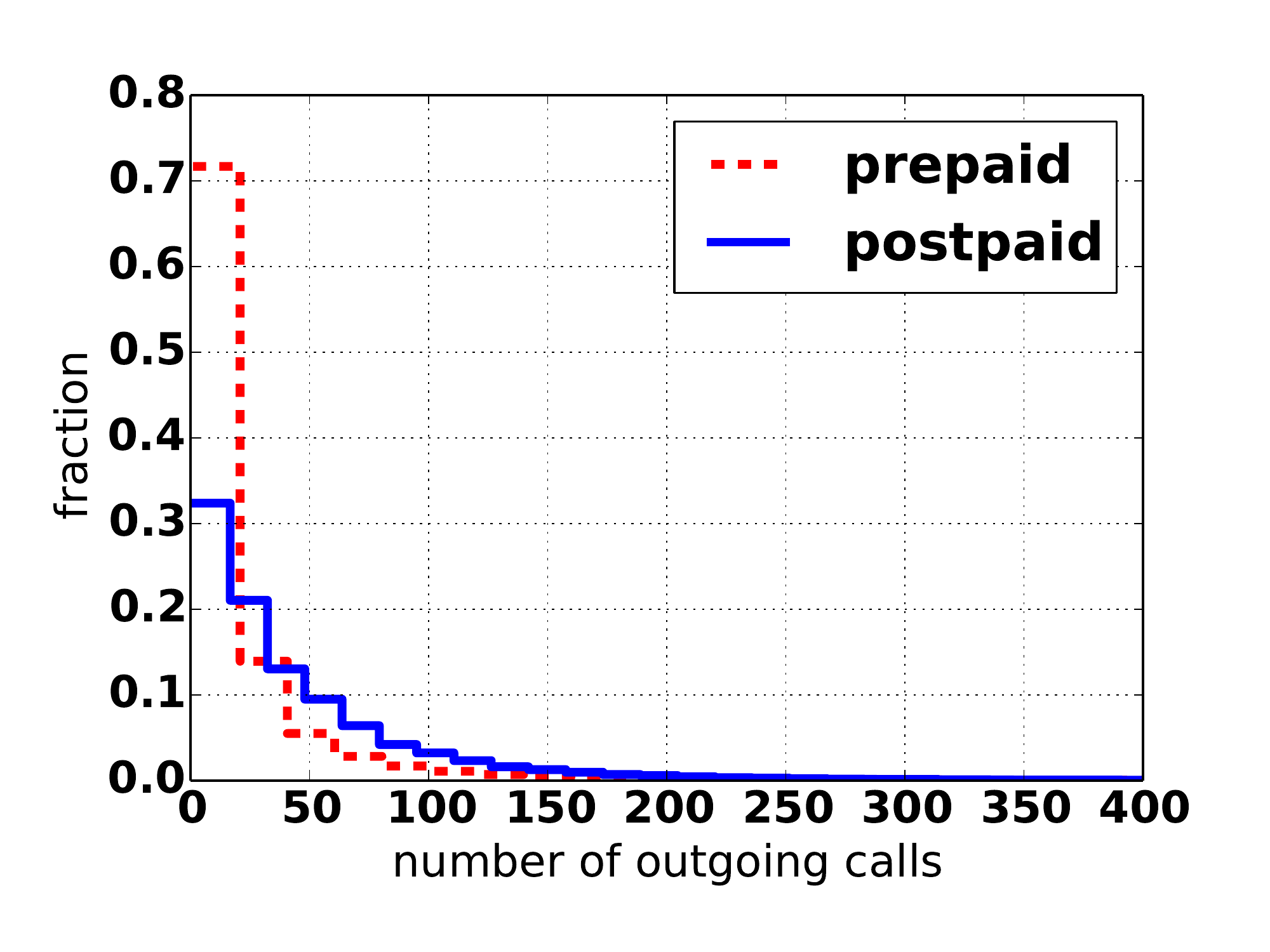}}
     \qquad
     \subfigure[]{\includegraphics[width=0.46\linewidth]{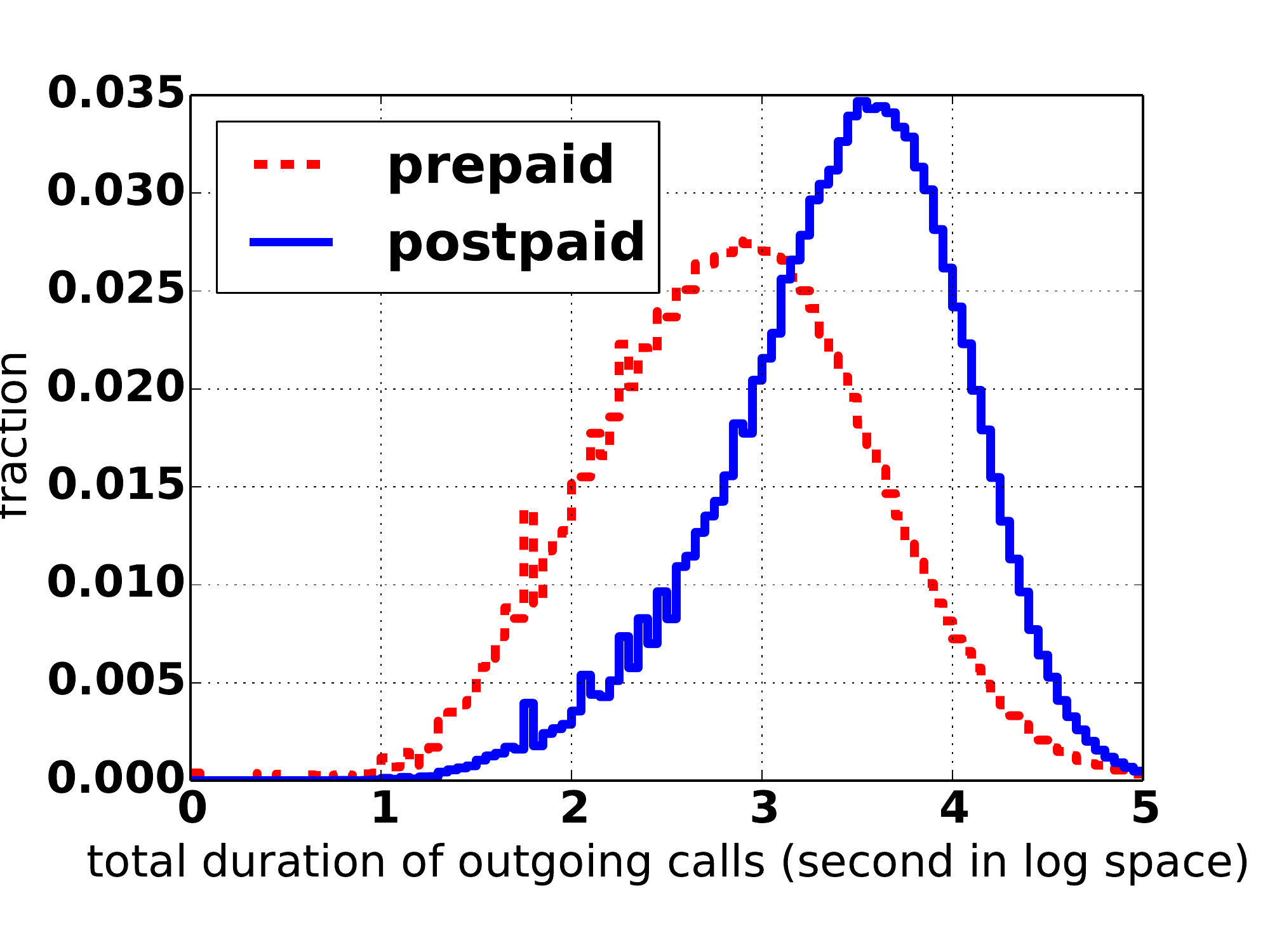}}\\
     \subfigure[]{\includegraphics[width=0.46\linewidth]{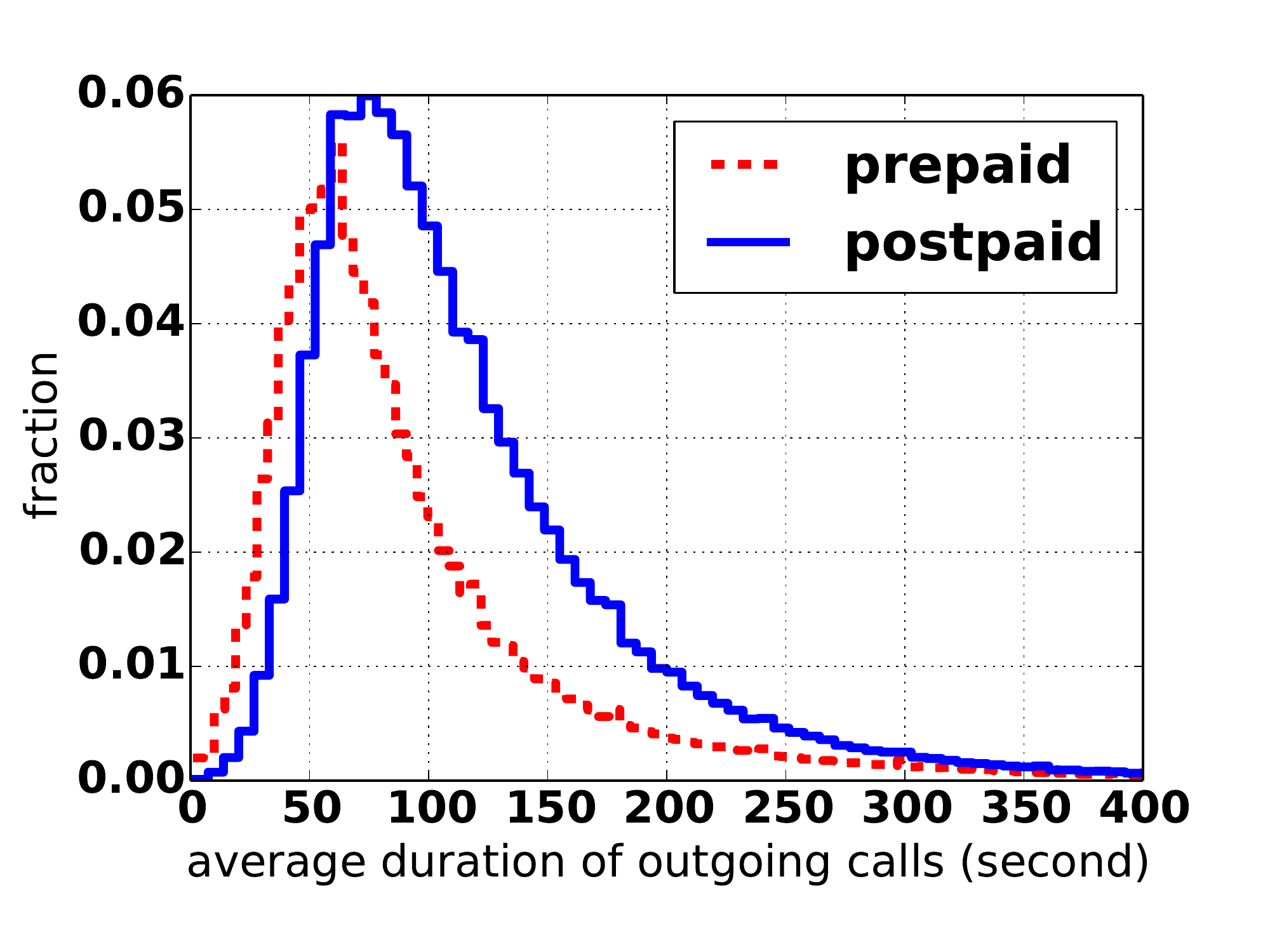}}
     \qquad
     \subfigure[]{\includegraphics[width=0.46\linewidth]{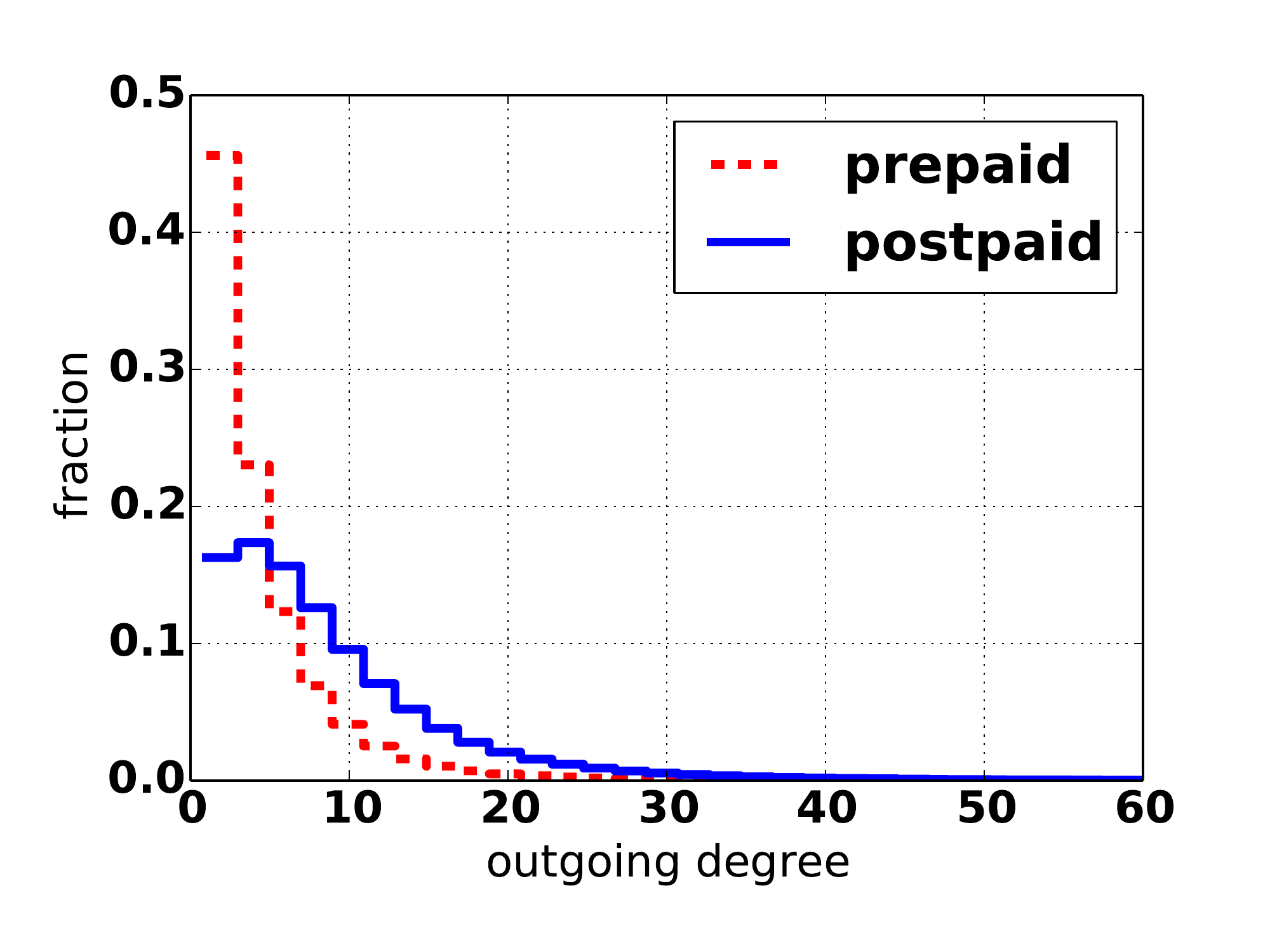}}
 \caption{From left to right are the distributions of each user for the total number of outgoing calls, the total (in the logarithmic space) and average duration of each outgoing call and the outgoing degree of each user. Note that these attributes are computed using the data of calls from the operator network of \A to that of B, whereas those in Figure \ref{fig:attributedistribution} are computed using the data of calls within the operator network of A.}
 \label{fig:attributedistribution_acrossnetwork}
\end{figure}

In our first method let's consider two mobile providers: \A represents our central company with available information about its company users; while \B is an external company (of non-company users) with no user information available. First we check whether characteristic call features, observed in Section \ref{sec:classification}, also differentiate between prepaid and postpaid users in \A but considering only inter-company communication. Here we take each users in \A from DS2, and using their CDRs only with users in \B, we measure the distributions of the number and duration of outgoing calls and their out-degree. As shown in Fig.\ref{fig:attributedistribution_acrossnetwork} these distributions yet appear to be very different for prepaid and postpaid \A users, just as it was shown earlier for intra-company communications. It suggests that these node attributes are not subjective to intra-company communications only but may generalise for any communication tie, even between users of different providers.

Based on these observations we assume that these features characterise symmetrically the communications between any users independent of their provider. However, since we have access to the information of \A users only, this is the only set we can use for direct verification of the results of an inference method. Thus in our first approach we took the point of the company \B, and we actually try to infer the subscription types of users in \A, by only using CDRs between the users of \A and \B and the characteristic  attributes shown in Fig.\ref{fig:attributedistribution_acrossnetwork}. We applied the same machine learning methods with the same setting as described in section \ref{sec:classification}, but now on the dataset DS2. This experiments showed that our method based merely on statistical node attributes alone can achieve an accuracy of $\sim 68.2\%$ using SVM and Boosting. The confusion matrix by Boosting is given in Table \ref{tab:confusion_classification_outside}. It demonstrates that using only inter-company communications and user attributes, one can infer the subscription types of customers of another provider with relatively high accuracy.

\begin{table}[h!]
\caption{Confusion matrix by Boosting.}
\centering
\begin{tabular}{|p{3.6cm}|p{3.6cm}|p{3.6cm}|}
\hline accuracy=\textbf{0.687}&{postpaid}&{prepaid}\\\hline
{postpaid}&\textbf{0.682}&0.318\\\hline
{prepaid}&0.317&\textbf{0.693}\\\hline
\end{tabular}
\label{tab:confusion_classification_outside}
\end{table}

\subsection{Inference using label propagation}

In our earlier data analysis, in Section \ref{sec:classification}, we found that calls between the same type of users (prepaid or postpaid) appeared $\sim 3$ times more frequently than between users holding different subscriptions. Although this was observed between only users of \A, we may assume similar homophilic correlations to be present between users of \A and \B. Based on this assumption here we present a second method to infer subscription types of non-company users using label propagation together with an indirect validation method.

Let's consider a bipartite network $G=(V_A,V_B,E_{AB})$, where $V_A$ is the set of all \A company users, $V_B$ is the set of all \B non-company users, and $E_{AB}$ contains all directed links between users in $V_A$ and $V_B$. A directed link is defined between $a\in V_A$ and $b\in V_B$ (resp. between $b$ and $a$) if $a$ called $b$ (resp. $b$ called $a$) at least once during the observation. In this network we know the subscription types of all users in $V_A$, while we would like to infer the subscriptions of $V_B$ users. As edges in $E_{AB}$ are real social ties inferred from communication records, they are not independent but show degree correlations, could be strong ties of communities, and may show homophilic correlations as we mentioned earlier. The knowledge of subscription types on side \A, together with the strong effects of subscription homophily, and the structurally correlated bipartite network provides us a potential way to define a label propagation method using majority rule to infer the subscription types of non-company users in $V_B$.

\begin{table}[h!]
\caption{Confusion matrix of label inference on the empirical structure.}
\centering
\begin{tabular}{|p{3.6cm}|p{3.6cm}|p{3.6cm}|}
\hline accuracy=\textbf{0.964}&{postpaid}&{prepaid}\\\hline
{postpaid}&\textbf{0.961}&0.039\\\hline
{prepaid}&0.034&\textbf{0.966}\\\hline
\end{tabular}
\label{tab:confusion_classification_orig}
\end{table}

More precisely, we take all users from $V_B$ and their incoming ties, and assign them a label (prepaid or postpaid) corresponding to the majority of the subscription type of their connected friends in $V_A$. In case of equality we selected randomly a label by respecting the overall balance between pre- and postpaid users observed in \A. However, even we can assign labels for all the users in $V_B$, we cannot validate the precision of our inference as we do not have access to the subscription types of non-company users. On the other hand, assuming all correlations to be symmetric between the two sets of users, we can use the inferred labels of $V_B$ users to assign labels to users in $V_A$ applying the same majority rule inference method. After these two ways of label inference we can validate the precision of the inferred labels of users in $V_A$ by comparing them to their real subscription types. The result will give us a lower bound for the inference error in $V_B$ as erroneous inference propagates also during the process, and the entropy of inference cannot increase during the two ways of inference. As a result we found that on average our method identified the correct labels of $96.4\%$ of users in $V_A$. Our measurements were averaged over $100$ independent realisations with results summarised in the confusion matrix shown in Table \ref{tab:confusion_classification_orig}.

This method could give trivial results if two conditions are satisfied at the same time: the average degree in the bipartite network is $1$, and all links are bidirectional. In our case, although the $89\%$ of links found to be bidirectional, the average degree is relatively high $\langle k \rangle\simeq 2.593$ suggesting that the inferred labels are not trivial. This can be further validated by eliminating structural correlations from the bipartite network and see whether we can recover the original results. If the accuracy of inference remains the same in the uncorrelated network, our results are trivial and structural correlations do not play a role here. On the other hand if the accuracy drops in the uncorrelated networks, our results are meaningful. To remove structural correlations we used a configuration network model scheme \cite{Newman:network}, i.e., we choose random pairs of links of the same direction and swapped them while disallowing double links. Repeating this step several times we received a random bipartite structure where the in- and out-degrees of nodes did not change but any structural correlation vanished from the network. Repeating our two-way label inference method on $100$ of such independently generated random bipartite structures we found that the precision dropped to $56.2\%$, with a standard deviation of $5\%$, which suggests that our original findings were significant. For precise results see the confusion matrix in Table \ref{tab:confusion_classification_randnet}.

\begin{table}[h!]
\caption{Confusion matrix of label inference on the randomised structure.}
\centering
\begin{tabular}{|p{3.6cm}|p{3.6cm}|p{3.6cm}|}
\hline accuracy=\textbf{0.562}&{postpaid}&{prepaid}\\\hline
{postpaid}&\textbf{0.539}&0.461\\\hline
{prepaid}&0.415&\textbf{0.585}\\\hline
\end{tabular}
\label{tab:confusion_classification_randnet}
\end{table}

\section{Conclusions and Future Work}
\label{sec:conclusions}

In this paper we presented novel methods to detect prepaid and postpaid customers in mobile phone datasets by exploiting both user attributes and observed structural correlations in the communication network. First we addressed the problem with known inference methods relying on subscription type labels known for some users. We showed that in this case SVM provides us the best solution with 0.891 accuracy.

Second, as the main contribution of our work, we addressed a more challenging problem, to infer subscription labels without prior knowledge of subscription types, but only relying on user attributes and observed structural correlations. We cast the classification problem as a problem of graph labelling and solved it by max-flow min-cut algorithms. With this novel methodology we achieved a classification accuracy of $\sim 87\%$ which is about $\sim 7\%$ better than supervised learning on user attributes alone.

Third, we provided two methods to infer subscription types of non-company users. We showed that using node attributes alone we can achieve already $\sim 70\%$ accuracy of inference. In addition, exploiting present homophilic structural correlations, we obtained good accuracy of inference by using a two-ways label propagation method.

Our results have certain limitations. As we explained, the incomplete knowledge of subscription types of non-company users makes difficult the full verification of any inference problems. We provided one possible indirect solution here but this should be developed further using more complete datasets. We also generalised some observations on node features and structural correlations for any node and link in the communication network, which should be verified. In addition, the two ways inference method provides more accurate results if the average out-degree of non-company users is high. For some reasons we found its value relatively small, which limits the precision of our inference results. We also note that indirect inference method works in case of competing provides with similar subscription portfolio and user base, thus where our observations on company users the best generalise.
 
In the future, there are several avenues to explore. First, one could extend our approach from binary classification to multi-class classification, which can be similarly formulated as a graph labelling problem but with more than two label nodes. Such problems are known to be NP-hard but greedy algorithms based on max-flow min-cut have been proposed to achieve approximate solutions~\cite{graphcuts}. Second, our dataset is intrinsically dynamical and contains CDRs over one month, what we all used to predicting user subscription types. It would be interesting to study the impacts of the dynamics of the mobile call data over time for longer periods to understand how temporal patterns of communications can improve our predictions. Possible structural biases in the network between users of different customers could be also a direction to explore. Our aim here was to contribute to the more general discussion of user attribute inference and to give possible directions for potential future applications in behavioural prediction and personal marketing.

\subsection*{Acknowledgment}
We acknowledge Grandata to share the data and M. Fixman for his technical support. This research project was partially granted by the SticAmSud UCOOL project, INRIA, SoSweet (ANR-15-CE38-0011-01) and CODDDE (ANR-13-CORD-0017-01).

\subsection*{Authors' contribution}
All authors read and verified the manuscript. YL, WD, MK, CS and EF participated in the writing; YL, WD, MK and EF designed the research; and YL run all numerical calculations.

\bibliographystyle{splncs03}

\end{document}